\documentclass[aps,a4paper,10pt,twocolumn,nofootinbib,eqsecnum]{revtex4} 
\usepackage[T1]{fontenc}
\usepackage{pslatex}
\usepackage{datetime}
\usepackage{amsmath}
\usepackage{amsfonts}
\usepackage{amsfonts}
\usepackage{mathrsfs}
\usepackage[mathscr]{euscript}
\usepackage{mathtools}
\usepackage{fancyhdr}
\usepackage{colordvi}
\usepackage[hypertex=true]{hyperref}
\usepackage{epsfig}
\usepackage{color}
\usepackage{bm}
\usepackage{tensor}

\def\overstrike#1#2{{\setbox0\hbox{$#2$}\hbox to \wd0{\hss
    $#1$\hss}\kern-\wd0\box0}}

\newcommand{\Mat}[1]{\pmb{{#1}}} 
\newcommand{\Tensor}[1]{\mathsf{#1}}
\newcommand{\MatrixB}[1]{[#1]}             
\newcommand{\MatrixM}[1]{\left[#1\right]}  
\newcommand{\dualF}{{\ast\! F}} 
\newcommand{\dualchi}{{\ast\!\! \chi}} 

\newcommand{\Co}[1]{\MakeTextUppercase{#1}}

        \DeclareMathOperator{\sgn}{sgn}

        \DeclareMathOperator{\grad}{\nabla}

\newdateformat{yymmdddate}{\THEYEAR/\twodigit{\THEMONTH}/\twodigit{\THEDAY}}

\newcommand{\peephole}{periscope}

\begin{document}
\title{Transformation devices: carpets in space and space-time}
\author{Paul Kinsler}
\email{Dr.Paul.Kinsler@physics.org}
\author{Martin W. McCall}
\affiliation{
  Blackett Laboratory, Imperial College London,
  Prince Consort Road,
  London SW7 2AZ, 
  United Kingdom.
}

\lhead{\includegraphics[height=5mm,angle=0]{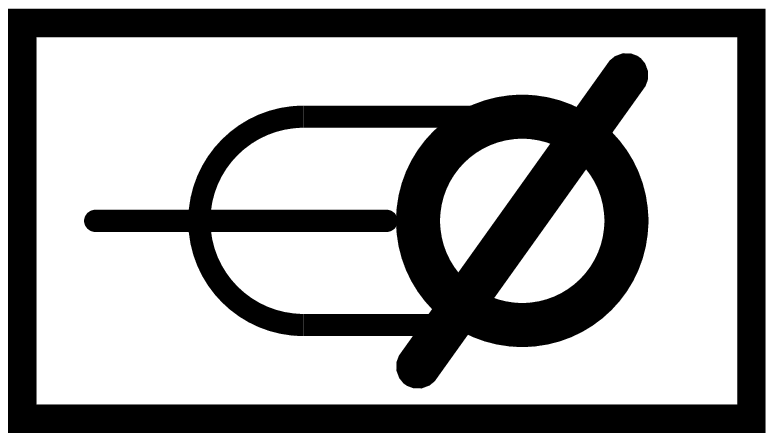}~~STCARPET}
\chead{T-devices: carpets in space \& time}
\rhead{
\href{mailto:Dr.Paul.Kinsler@physics.org}{Dr.Paul.Kinsler@physics.org}\\
\href{http://www.kinsler.org/physics/}{http://www.kinsler.org/physics/}
}

\begin{abstract}

Here we extend the theory of space-time or event cloaking
 into that based on the carpet or ground-plane reflective surface.
Further, 
 by recasting and generalizing a scalar acoustic wave model 
 into a new mathematically covariant form, 
 we also show how transformation theories for optics 
 and acoustics can be combined into a single prescription.
The single prescription, 
 however, 
 still respects the fundamental differences
 between electromagnetic and acoustic waves, 
 which then provide us with an existence  test for any
 desired transformation device --
 are the required material properties 
 (the required constitutive parameters) 
 physically permitted by the wave theory 
 for which it is being designed?
Whilst electromagnetism is a flexible theory
 permitting almost any transformation-device (T-device) design, 
 we show that the acoustic model used here is more restricted.

\end{abstract}


\date{\today}
\maketitle
\thispagestyle{fancy}

%

%
\section{Introduction}\label{S-intro}

The concept of electromagnetic cloaking has now 
 been with us for more than five years
 \cite{Pendry-SS-2006sci,Leonhardt-2006sci},
 but has been recently revitalized by the introduction 
 of the concept of space-time cloaking
 \cite{McCall-FKB-2011jo,Kinsler-M-2014adp-scast}.
Here, 
 in light of the many variants of spatial cloaking
 that now exist --
 ordinary cloaks, 
 carpet cloaks \cite{Li-P-2008prl}, 
 exterior cloaks \cite{Lai-CZC-2009prl}, 
 we extend the possible implementation
 to include space-time \emph{carpet} cloaks.
Of course, 
 cloaking is not the only application of transformation theories, 
 as evidenced not only by spatial illusion devices
 \cite{Lai-NCHXZC-2009prl,Zang-J-2011josab,Chen-C-2010jpd}, 
 but also ideas for space-time illusion devices \cite{Kinsler-M-2014adp-scast}
 that appear to trick --
 but only trick --
 causality \cite{Kinsler-2011ejp}.
Other applications
 such as beam control \cite{Heiblum-H-1975jqe,Ginis-TDSV-2012njp}
 geodesic lenses
 \cite{Sarbort-T-2012jo,Kinsler-TTTK-2012ejp}, 
 hyperbolic materials
 \cite{Smolyaninov-2013jo,HypMetaM-2013oe},
 and cosmological models
 \cite{Faraoni-GN-1999fcp,Kinsler-M-2014adp-scast,Chen-ML-2010oe,Smolyaninov-YBS-2013oe}
 are also possible.

Like a spatial carpet cloak, 
 the space-time ``event'' carpet cloak is one sided and reflective,
 and is less singular than a traditional spatial cloaking implementation.
Rather than hide a region of space in perpetuity, 
 however, 
 the ``event'' carpet cloak allows the region
 to always appear visible, 
 even though a finite segment of its timeline
 is forced to take place in darkness.
This dark period means that events therein 
 are edited out of visible history.
However, 
 those missing events are covered up by clever manipulation of the
 light signals that communicate to observers
 their information about events.
Alternatively, 
 the transformation can be adapted
 to temporarily include extra events, 
 rather than exclude them; 
 thus creating a space-time ``{\peephole}''.
We might even relax our preference for invisibility devices, 
 and construct one that appears bigger on the inside than the outside -- 
 a ``tardis''.

In this paper we deliberately use \emph{first order} equations
 to model our wave mechanics \cite{Geroch-1996-PDE}, 
 so that a pair of them are needed 
 (in concert with constitutive or state equations of some kind)
 to generate wave behaviour.
If desired, 
 the first order equations can be substituted inside
 one another to give a familiar second order form, 
 but in fact the first order formulation is both less restrictive
 and more general.
Indeed, 
 we will see that it is suggestive of a generalization 
 to a \emph{Transformation Mechanics}
 valid for any wave equations expressible in such a form, 
 being also related to a ``transformation media'' concept
 addressing the material properties required by T-devices.
Notably, 
 here we apply the event carpet cloak and {\peephole} transformations
 to both electromagnetism (EM)
 and a simple pressure-acoustics (p-acoustics) model, 
 demonstrating how specific wave models allow or restrict
 the set of possible T-devices.

We introduce the ``wave mechanics'' models
 for EM and our p-acoustics
 in section \ref{S-waves}, 
 and show how the two can be unified.
This then allows us to unify 
 transformation optics,
 transformation p-acoustics, 
 and indeed any other compatible wave transformation theory
 into a general transformation mechanics.
After this, 
 we specialize to ground-plane T-devices --
 in section \ref{S-carpet} we investigate those 
 based on the (spatial) carpet cloak, 
 and in \ref{S-stcarpet} we generalize them into 
 space-time carpet T-devices.
In section \ref{S-implementation}
 we explain the operation of space-time carpet cloaks,
 and propose a scheme for implementing one for EM waves,
 before concluding in section \ref{S-summary}.

%
\section{Waves and transformations}\label{S-waves}

In this paper we do not restrict ourselves to
 a single type of wave mechanics in which to construct 
 ground plane T-devices.
Instead,  
 we will show that a unified procedure is possible, 
 encompassing both electromagnetism 
 and a simple pressure-acoustics model; 
 and indeed any linear wave mechanics that can be described
 within a theory containing two second-rank field tensors and 
 one fourth-rank constitutive tensor.
We will start with the ordinary vector calculus descriptions, 
 and then show how they can be re-expressed in the tensor form; 
 and also describe how the tensor descriptions can 
 be abbreviated into a more accessible matrix form.
This process is of course well known for EM, 
 but here we use an unconventional tensorization
 so that it more easily matches up with the p-acoustic description 
 and other possible generalisations, 
 as well as with straightforward matrix algebra versions.
A simpler 1D introduction to this description
 can be seen in \cite{Kinsler-M-2014adp-scast}.
After the wave theory descriptions, 
 we show how transformations intended
 to implement specific T-devices
 affect the differential equations and constitutive relations.

%
\subsection{Electromagnetism}\label{SS-waves-optics}

%

Maxwell's equations \cite{Jackson-ClassicalED} are
 often written as a set of vector differential equations
 with two pairs of field types,
 and two constitutive relations.
These are
~
\begin{align}
  \partial_t D(\Vec{r},t)
&=
  \grad \times \Vec{H}(\Vec{r},t)
 -
  \Vec{J}_f(\Vec{r},t)
\label{eqn-Maxwell-dt-D}
\\
\quad 
  \grad \cdot \Vec{D}
&=
  \rho_f
;
\label{eqn-Maxwell-divD}
\\
  \partial_t B(\Vec{r},t)
&=
 -
  \grad \times \Vec{E}(\Vec{r},t)
\label{eqn-Maxwell-dt-B}
,
\\
  \grad \cdot \Vec{B}
&=
  0,
\label{eqn-Maxwell-divB}
\end{align}
 with constitutive relations 
~
\begin{align}
  \Vec{B} = \Mat{\mu}\Vec{H} - \Mat{\mu} \Mat{\beta}\Vec{E} 
 &\quad \rightarrow \quad
  \Vec{H} = \Mat{\eta}\Vec{B} + \Mat{\beta}\Vec{E} 
,
\label{eqn-Maxwell-constit-HB}
\\
  \Vec{D} = \Mat{\epsilon}' \Vec{E} + \Mat{\mu} \Mat{\alpha} \Vec{H}
 &\quad \rightarrow \quad
  \Vec{D} = \Mat{\epsilon} \Vec{E} + \Mat{\alpha} \Vec{B}
\label{eqn-Maxwell-constit-DE}
,
\end{align}
 where the permittivity matrix is 
  $\Mat{\epsilon}' = \Mat{\epsilon} - \Mat{\alpha} \Mat{\mu} \Mat{\beta}$, 
 and $\Mat{\eta}$ is the inverse of the 
 more common permeability matrix $\Mat{\mu}$.
Although in many circumstances the left hand (LH) form
 is preferred, 
 for the tensor form discussed next the right hand (RH) form is used instead.
The matrices $\Mat{\alpha}$ and $\Mat{\beta}$ 
 contain information about magnetoelectric coupling, 
 and follow $\Mat{\beta} = \Mat{\alpha}^\dagger$ \cite{Post-FSEM}.

Since we are interested primarily in freely propagating EM fields, 
 we will assume that 
 that there are no free electric charges 
 (i.e. $\rho_f=0$), 
 and as is usual,
 we have already assumed in eqn. \eqref{eqn-Maxwell-divB}
 that there are no magnetic charges or currents.

%

In tensor form the differential equations 
 have a remarkably simple form.
In order to emphasize their similarities, 
 we chose to use the usual $\Tensor{G}$ tensor density 
 (i.e. ${G}\indices{^{\alpha \beta}}$)
 but the dual of the $\Tensor{F}$ tensor 
 (i.e. $\Tensor{\dualF}$,
 with components 
 $\dualF\indices{^{\alpha \beta}} 
  = \frac{1}{2} F\indices{_{\mu\nu}} 
        \varepsilon\indices{^{\mu\nu\alpha \beta}}$), 
 where $\varepsilon\indices{^{\mu\nu\alpha \beta}}$
  is the antisymmetrization symbol.
The equations can then be written as
~
\begin{align}
  \partial\indices{_{\alpha}}
  \dualF\indices{^{\alpha \beta}}
=
  0
,
\qquad
  \partial\indices{_\beta}
  G\indices{^{\alpha \beta}}
=
  J\indices{^{\alpha}}
.
\label{eqn-Maxwell-tensorial}
\end{align}
Using $\Tensor{G}$ and $\Tensor{\dualF}$ means that
 we must then use
 a mixed form for the dual $\Tensor{\dualchi}$
 of the constitutive tensor $\Tensor{\chi}$ 
 in the constitutive relations, 
 with
~
\begin{align}
  G\indices{^{\alpha \beta}} 
&=
  \frac{1}{2} 
  \dualchi\indices{^{\alpha\beta}_{\gamma\delta}}
  \dualF\indices{^{\gamma\delta}}
.
\end{align}
This use of $\dualF\indices{^{\alpha \beta}}$, 
 $G\indices{^{\alpha \beta}}$, 
 and $\dualchi\indices{^{\alpha\beta}_{\gamma\delta}}$,
 both generalizes better
 and
 maps onto the vector calculus representation more simply.

Note that since $J\indices{^{\alpha}}$
 (and its vector counterpart $\Vec{J}$)
 is a current density, 
 this means that $\Tensor{G}$ is also a density.
Likewise, 
 although the usual EM tensor $\Tensor{F}$ is not a density, 
 its dual $\Tensor{\dualF}$ is.
Thus 
 eqns. \eqref{eqn-Maxwell-tensorial}
 must be transformed as densities.
Consequently, 
 the mixed-form
 constitutive tensor $\dualchi\indices{^{\alpha\beta}_{\gamma\delta}}$
 is \emph{not} a density.

In terms of generating wave solutions, 
 the constitutive relations
 connect the two tensor differential equations together
 into a single system, 
 and so allow (wave) solutions to be found (i.e. exist).
This is most easily seen using the vector form, 
 where we can use the constitutive relations to allow us to substitute
 one of the Maxwell curl equations into the other, 
 giving us one of the four \cite{Kinsler-FM-2009ejp} 
 possible electromagnetic second order wave equations.

Matrix representations of the antisymmetric
 $\Tensor{\dualF}$ and $\Tensor{G}$
 can be written out in $\MatrixB{t,x,y,z}$ coordinates, 
 as
~
\begin{align}
  \MatrixM{ \dualF\indices{^{\alpha\beta}} }
&=
    \begin{bmatrix}
       0   &    B_x  &    B_y  &    B_z \\
     -B_x  &     0   & -  E_z  &    E_y \\
     -B_y  &    E_z  &     0   & -  E_x \\
     -B_z  & -  E_y  &    E_x  &     0 
    \end{bmatrix}
,
\\
  \MatrixM{ G\indices{^{\alpha\beta}} }
&=
    \begin{bmatrix}
       0   &    D_x  &    D_y  &    D_z \\
     -D_x  &     0   &    H_z  & -  H_y \\
     -D_y  & -  H_z  &     0   &    H_x \\
     -D_z  &    H_y  & -  H_x  &     0 
    \end{bmatrix}
.
\end{align}
Here, 
 the $\partial\indices{_\alpha}$ has become
 the row vector 
 $\MatrixB{ \partial_t, \partial_x, \partial_y, \partial_z }$;
 and the current density $J\indices{^\alpha}$ also becomes a row vector.

The constitutive tensor $\dualchi\indices{^{\alpha\beta}_{\gamma\delta}}$
 is rank four so that a direct matrix representation is too big
 to display.
Nevertheless, its 
 permittivity entries convert $E_i$ from $\dualF\indices{^{\gamma\delta}}$ 
 into
 $D_i$ from $G\indices{^{\alpha \beta}}$, 
 thus they will have the same units as permitivitty
 (i.e. as $\epsilon$); 
 likewise
 the permeability entries convert $B_i$ from $\dualF\indices{^{\gamma\delta}}$
 into
 $H_i$ from $G\indices{^{\alpha \beta}}$, 
 thus they will have the same units as inverse permeability
 (i.e. as $\mu^{-1} = \eta$).

%

We can follow the usual method of achieving a convenient matrix-like 
 compacted representation for $\dualF\indices{^{\alpha \beta}}$
 and $G\indices{^{\alpha \beta}}$, 
 using the fact that each tensor has only 6 unique entries.
For all combinations of indices, 
 we select out the relevant row and column coordinate pairs
 $tx, ty, tz, yz, zx, xy$ 
 in sequence, 
 so we can write $\dualF\indices{^{\alpha \beta}}$
 and $G\indices{^{\alpha \beta}}$
 as column vectors $\MatrixB{\dualF\indices{^A}}$ and $\MatrixB{G\indices{^B}}$ in turn, 
 being
~
\begin{align}
  \MatrixM{\dualF\indices{^A}}
&=
  \MatrixM{ ~B_x, ~B_y, ~B_z,-E_x,-E_y,-E_z}^T
, 
\\
  \MatrixM{G\indices{^B}} 
&=
  \MatrixM{ ~D_x, ~D_y, ~D_z, ~H_x, ~H_y, ~H_z}^T
.
\end{align}
As a result the rank-4 constitutive tensor 
 can also be compacted, 
 and so be represented by a $6 \times 6$ matrix.
Again the first (upper) index $A$ spans the matrix rows,
 so that
~
\begin{align}
  \MatrixM{ \dualchi\indices{^{A}_{B}} }
&=
  \left[
    \begin{array}{c|c}
      ~\Mat{\alpha}     & -\Mat{\epsilon} \\
     \textrm{----}     & \textrm{----}  \\
      ~\Mat{\eta}       & -\Mat{\beta}    \\
    \end{array}
  \right]
,
\label{eqn-EM-referencechi}
\end{align}
 so that the constitutive relation now has a matrix form, 
~
\begin{align}
  \MatrixM{ G\indices{^A} }
&=
  \MatrixM{ \dualchi\indices{^{A}_{B}} }
  \MatrixM{ \dualF\indices{^B} }
,
\end{align}
 which is particularly convenient for manual calculation.
Normally when using this kind of representation,
 the all-upper index tensor $\dualchi\indices{^{\alpha\beta\gamma\delta}}$
 is used, 
 but here we are using the mixed tensor form
 $\dualchi\indices{^{\alpha\beta}_{\gamma\delta}}$.
This means that our matrix expression looks different; 
 notably a simple isotropic non-magneto-electric medium, 
 with a diagonal $\dualchi\indices{^{\alpha\beta\gamma\delta}}$ (matrix)
 is now block off-diagonal.

Note that 
 the matrix $\MatrixB{\dualchi}$ only includes half
 of all the allowed elements of $\dualchi\indices{^{\alpha\beta}_{\gamma\delta}}$, 
 but the missing elements are all duplicates of included
 ones\footnote{This is why no 1/2 appears
  in the \emph{matrix} constitutive relations.
 However, 
  we must still account for this doubling up
  when evaluating transformations later.}.
However, 
 each element of $\MatrixB{\dualchi}$ nevertheless independently links
 any component of $\MatrixB{F}$ to any other of $\MatrixB{G}$; 
 there is no duplication, 
 or \emph{a priori} requirement that certain constitutive parameters 
 must be equal or otherwise closely related.
As a result, 
 each constitutive property of an electromagnetic material
 can (at least mathematically) be adjusted independently,
 meaning that any well specified T-device might be constructed.

%
\subsection{Pressure Acoustics}\label{SS-waves-acoustic}

Here we introduce a simplified 
 pressure acoustic (p-acoustic) theory, 
 equivalent to one 
 describing linear (perturbative) acoustic waves
 on a stationary background fluid; 
 although it also encompasses many other types of wave  
 which contain a scalar component.
For linearized acoustic waves \cite{MorseIngard-Acoustics}
 in pressure $p$ and fluid velocity $\Vec{u}$, 
 we have constitutive parameters 
 $\bar{\kappa}$ (compressibility)
 and $\bar{\rho}$ (mass density).    
As well as the traditional quantities, 
 we also specify both a scalar $\Co{\bar{p}}$, 
 and a momentum density $\Vec{\Co{v}}$, 
 giving a total of four quantities that in combination 
 emphasize both the similarities and differences 
 compared with the description of EM; 
 this can be compared to e.g. the treatment by Sklan \cite{Sklan-2010pre}.
It also means that 
 p-acoustics is also a more general model than the comparable
 traditional ones, 
 which typically reduce the equations back down to 
 $p$ and $\Vec{u}$ --
 or even just a second order wave equation in $p$.

Here, 
 however, 
 in order to maximize the similarity with the EM notation, 
 we will substitute the scalar $\Co{\bar{p}}$
 with a number density $\Co{p}$, 
 the fluid velocity $\Vec{u}$ 
 with a velocity density $\Vec{v}$,
 and the constitutive parameters become 
 an inverse energy $\kappa$
 and a particle mass $\rho$.
Since the background fluid is stationary, 
 for small amplitude waves the $(\Vec{v} \cdot \grad) \Vec{v}$ term
 that would usually appear in acoustic wave equations 
 is second order, 
 and hence can be neglected.

%

For scalar quantities $\Co{p}$ and $p$, 
 and vectors $\Vec{\Co{v}}$ and $\Vec{v}$, 
 the wave equation pair, 
 here used to model p-acoustics, 
 is
~
\begin{align}
  \partial_t \Co{p} (\Vec{r},t)
&=
 -
  \grad 
  \cdot
  \Vec{v}(\Vec{r},t)
 +
  Q_{\Co{p}}(\Vec{r},t)
,
\label{eqn-wave-basic-p}
\\
  \partial_t \Vec{\Co{v}} (\Vec{r},t)
&=
 -
  \grad 
  p(\Vec{r},t)
 +
  \Vec{Q}_{\Co{v}}(\Vec{r},t)
,
\label{eqn-wave-basic-P}
\\
  \grad \times \Vec{\Co{v}} &= \Vec{\Sigma}, 
\qquad \qquad
  \grad \times \Vec{v} = \Vec{\sigma}
,
\label{eqn-wave-basic-src}
\end{align}
 where in simple cases
 $\Co{p} = \kappa p$ is a number density related to the pressure $p$
 by an inverse energy $\kappa$, 
 and the momentum density $\Vec{\Co{v}} = \Mat{\rho} \Vec{v}$
 is related to the velocity-field density $\Vec{v}$
 by a matrix of mass parameters $\Mat{\rho} = \MatrixB{\rho\indices{^i_j}}$.
There are two allowed types of source, 
 a particle number (density) source $Q_{\Co{p}}$, 
 and a momentum (density) source $\Vec{Q}_{\Co{v}}$.

The differential equation eqn. \eqref{eqn-wave-basic-p}
 is related to the conservation of particle number 
 (and conservation of mass)
 in a microscopic acoustic model, 
 and eqn. \eqref{eqn-wave-basic-P} 
 ensures conservation of momentum.
The third equation shows the p-acoustics analogue
 of charge, 
 and just as for the EM case, 
 where we are interested primarily in freely propagating fields,
 we set these to zero
 ($\Vec{\Sigma}=0, ~\Vec{\sigma}=0$).

The most general constitutive relations for 
 coupling between the scalar and vector fields, 
 can be written
~
\begin{align}
  \Co{p}       = \kappa^{-1} p - \kappa^{-1}\Vec{\alpha} \cdot \Vec{v}
  &\quad \rightarrow \quad
  {p}       = \kappa \Co{p} + \Vec{\alpha} \cdot \Vec{v}
\label{eqn-pacoustic-constit-Pp}
\\
  \Vec{\Co{v}} = \Vec{\beta} \kappa^{-1} p + \Mat{\rho}' \cdot \Vec{v}
  &\quad \rightarrow \quad
  \Vec{\Co{v}} = \Vec{\beta} P + \Mat{\rho} \cdot \Vec{v}
\label{eqn-pacoustic-constit-Vv}
,
\end{align}
 where 
 $\Mat{\rho} \cdot \Vec{v} 
  = \Mat{\rho}' \cdot \Vec{v} 
    + \Vec{\beta} (\Vec{\alpha} \cdot \Vec{v})/\kappa$.
Although in many circumstances the first (LH) form
 might be used,
 for the tensor form discussed next the second (RH) form is instead preferred.
The vector $\Vec{\alpha}$ 
 parameterizes some kind of 
 a velocity $\rightarrow$ pressure coupling, 
 and $\Vec{\beta}$ parameterizes
 a pressure $\rightarrow$ momentum density coupling.

Here the physical meanings of $\Vec{v}, ~\Vec{\Co{v}}$, 
 and $p, ~\Co{p}$
 mean that both eqn. \eqref{eqn-wave-basic-p} 
 \eqref{eqn-wave-basic-P} 
 transform like densities, 
 but the constitutive relations 
 eqns. \eqref{eqn-pacoustic-constit-Pp}, \eqref{eqn-pacoustic-constit-Vv}
 do not --
 just as for EM.

%

Just as for EM, 
 we can now embed $\Co{p}, \Vec{v}$
 into a tensor density, 
 which to ensure compatible notation we will call
 $\Tensor{\dualF}$, 
 with contravariant components  $\dualF\indices{^{\alpha \beta}}$.
We also embed the density fields 
 $p, \Vec{\Co{v}}$ into a tensor density $\Tensor{G}$
 with components $G\indices{^{\alpha \beta}}$; 
 and populate a constitutive tensor $\dualchi$ 
 with $\kappa, \Mat{\rho}, \Vec{\alpha}, \Vec{\beta}$
 appropriately.
The p-acoustic differential equations now appear
 in the same form as for EM, 
 being
~
\begin{align}
  \partial\indices{_\alpha}
  \dualF\indices{^{\alpha \beta}}
=
  J\indices{^{\beta}}
,
\qquad
  \partial\indices{_\beta}
  G\indices{^{\alpha \beta}}
=
  K\indices{^{\alpha}}
,
\end{align}
 with source terms $J\indices{^\alpha}, K\indices{^\alpha}$.
The constitutive relations are
~
\begin{align}
  G\indices{^{\alpha \beta}}
&=
  \dualchi\indices{^{\alpha\beta}_{\gamma\delta}}
  \dualF\indices{^{\gamma\delta}}
.
\end{align}
Here the only differences from EM are
 the internal structure -- 
 how the tensors are populated -- 
 and no factor of one half in the constitutive relations.
Again, 
 the $\Tensor{G}$ and $\Tensor{\dualF}$ tensor densities
 and their differential equation must transform like densities,
 but the constitutive tensor $\Tensor{\dualchi}$ does not.

In $\MatrixB{t, x, y, z}$ coordinates 
 the tensor $\dualF\indices{^{\alpha \beta}}$
 and
 density tensor $G\indices{^{\alpha \beta}}$ 
 can be written out
 using field matrices
~
\begin{align}
  \MatrixM{ \dualF\indices{^{\alpha \beta}} }
&=
    \begin{bmatrix}
   \Co{p}  &   0   &   0   &   0  \\
    ~ v_x  &   0   &   0   &   0  \\
    ~ v_y  &   0   &   0   &   0  \\
    ~ v_z  &   0   &   0   &   0  \\
    \end{bmatrix}
,
\quad
  \MatrixM{ G\indices{^{\alpha \beta}} }
&=
    \begin{bmatrix}
       0   &    0      &    0      &    0     \\
 ~\Co{v}_x &    p      &    0      &    0     \\
 ~\Co{v}_y &    0      &    p      &    0     \\
 ~\Co{v}_z &    0      &    0      &    p     \\
    \end{bmatrix}
,
\end{align}
 with
 source terms $J^0 = Q_{\Co{p}}$
 and $\Vec{K} = \MatrixB{K^i} = \Vec{Q}_{\Co{v}}$.
However,
 both $\Vec{J} = \MatrixB{J^i} = 0$
 and $K^0 = 0$.

Here the constitutive tensor $\dualchi\indices{^{\alpha\beta}_{\gamma\delta}}$
 relates the $P$ element from $\dualF\indices{^{\gamma\delta}}$
 to the $p$ from $G\indices{^{\alpha \beta}}$,
 and is thus $\sim \kappa$; 
 likewise the $v_i$ elements from $\dualF\indices{^{\gamma\delta}}$
 are related to $\Co{v}_i$ from $G\indices{^{\alpha \beta}}$
 by $\sim \rho$.

Note, 
 however, 
 that while this conveniently encodes the model of pressure acoustics
 we consider here, 
 it is not the most general that might be formulated.
The most general theory would be to consider field tensors
 that are symmetric, 
 to contrast with the EM representation using antisymmetric field tensors.
This would give not two sets of four amplitudes
 (i.e. $P$ and $v_i$, with $p$ and $V_i$), 
 but two sets of 10 amplitudes.
However, 
 we leave discussion of such details to a later work; 
 including describing how pentamode acoustics \cite{Norris-2008rspa} 
 can be represented in this way.

%

As with EM, 
 the tensor constitutive representation can be compacted.
 Here we use the fact that the elements $G^{xx}$, $G^{yy}$ and $G^{zz}$
 are all just $p$, 
 and so write a compact vector for $\Tensor{G}$ using this knowledge; 
 similarly for $\Tensor{\dualF}$.
The 
 resulting field column vectors
 $\MatrixB{G\indices{^A}}$, $\MatrixB{\dualF\indices{^B}}$
 are 
~
\begin{align}
  \MatrixM{ G\indices{^A} }
&=
  \MatrixM{ p, \Co{v}_x, \Co{v}_y, \Co{v}_z }^T
\label{eqn-pAcou-referenceG-short}
\\ 
  \MatrixM{ \dualF\indices{^B} }
&=
  \MatrixM{ \Co{p}, v_x, v_y, v_z }^T
.
\label{eqn-pAcou-referenceF-again}
\end{align}
The matrix representation of $\dualchi\indices{^{\alpha\beta}_{\gamma\delta}}$
 has the upper index denoting rows, 
 and the lower index denoting columns, 
 so that the $4 \times 4$ constitutive matrix is
~
\begin{align}
  \MatrixM{ \dualchi\indices{^{\alpha\beta}_{\gamma\delta}} }
=
  \MatrixM{ \dualchi\indices{^{A}_{B}} }
&=
  \left[
    \begin{array}{c|c}
         \kappa      & \Vec{\alpha}^T \\
      \textrm{------}  & \textrm{------}      \\
      \Vec{\beta}  &  \Mat{\rho}        \\
    \end{array}
  \right]
\label{eqn-pAcou-referencechi-short}
\end{align}
 the matrix constitutive relation is 
~
\begin{align}
  \MatrixM{ G\indices{^A} }
&=
  \MatrixM{ \dualchi\indices{^{A}_{B}} } 
  \MatrixM{ \dualF\indices{^B} }
\\
    \begin{bmatrix}
      G^{ww} \\
      G^{xt} \\
      G^{yt} \\
      G^{zt} 
    \end{bmatrix}
&=
    \begin{bmatrix}
     ~\kappa     & ~\alpha_x   & ~\alpha_y   & ~\alpha_z     \\
     ~\beta_x    & ~\rho_{xx}  & ~\rho_{xy}  & ~\rho_{xz} \\
     ~\beta_x    & ~\rho_{yx}  & ~\rho_{yy}  & ~\rho_{yz} \\
     ~\beta_x    & ~\rho_{zx}  & ~\rho_{zy}  & ~\rho_{zz}
    \end{bmatrix}
    \begin{bmatrix}
      \dualF^{tt} \\
      \dualF^{xt} \\
      \dualF^{yt} \\
      \dualF^{zt} 
    \end{bmatrix}
.
\label{eqn-pAcou-referenceConstEq-short}
\end{align}
Note that we \emph{must} remember that 
 the $G^{ww}$ element can be freely substituted
 by any of $G^{xx}$, $G^{yy}$ or $G^{zz}$; 
 this acts as an implicit constraint on allowed transformations, 
 which are only allowed to transform 
 $x$, $y$, and $z$ equivalently\footnote{A
  more explicit version of the $\MatrixB{G\indices{^A}}$ vector might 
  have six elements, 
  the first three of which would represent $G^{xx}$, $G^{yy}$ or $G^{zz}$
  in turn, 
  but all being equal to $p$; 
  and with $\MatrixB{\dualchi\indices{^{A}_{B}}}$ being a $6 \times 4$ matrix,
  the top three rows being identical.
  We must take care to allow for this multiplicity
  when transforming $\MatrixB{G\indices{^A}}$.}.
{In a scalar wave theory such as p-acoustics, 
 all non-isotropic \emph{wave} behaviour has to 
 derive from the constitutive parameters
 $\Vec{\alpha}$, $\Vec{\beta}$, or $\Mat{\rho}$, 
 and {not} from a transformation.}


%
\subsection{Transformations}\label{S-transformations}

With both of our preferred sorts of waves 
 having been expressed using the same mathematical machinery, 
 we can now investigate the effect of coordinate transformation
 in either case simultaneously, 
 leaving any more specific details to a later stage.
Deforming transformations 
 (or, technically, diffeomorphisms \cite{Nakahara-GTP})
 of the coordinates 
 (and hence of the matrix representations of the field
 and constitutive tensors)
 are simple to handle.
Transformations
 between the original coordinates $x^\alpha$
 and new ones $x^{\alpha'}$ 
 are easily written down as 
~
\begin{align}
  T\indices{^{{\alpha}'}_{{\alpha}}}
&=
  \MatrixM{
    \frac{\partial {x^{\alpha'}}}{\partial {x^\alpha}}
  }
,
\end{align}
 so that the field density tensors 
 $\Tensor{G}$ and $\Tensor{F}$ transform as
~
\begin{align}
  G\indices{^{\alpha'\beta'}}
&=
  {\Delta}^{-1}
  T\indices{^{{\alpha}'}_{{\alpha}}}
  T\indices{^{{\beta}'}_{{\beta}}}
  G\indices{^{\alpha\beta}}
\quad=
  L\indices{^{\alpha'}_{\alpha}^{\beta'}_{\beta}}
  G\indices{^{\alpha\beta}}
,
\\
  \dualF\indices{^{\gamma'\delta'}}
&=
  {\Delta}^{-1}
  T\indices{^{\gamma}_{\gamma'}}
  T\indices{^{\delta}_{\delta'}}
  \dualF\indices{^{\gamma\delta}}
\quad=
  M\indices{^{\gamma}_{\gamma'}^{\delta}_{\delta'}}
  \dualF\indices{^{\gamma\delta}}
.
\end{align}
The factor $\Delta = \left|\det (T)\right|$ occurs because 
 the fields are represented by tensor \emph{densities} 
 rather than a pure tensor.
Using the modulus of the determinant ensures that 
 parity transformations changing the handedness
 still preserve positive volumes.


Now we can address the transformation 
 of the constitutive tensor $\Tensor{\dualchi}$.
If for EM we set $a=2$, 
 and for p-Acoustics we set $a=1$, 
 then 
~
\begin{align}
  G\indices{^{\alpha'\beta'}}
&=
  \frac{1}{a}
  \dualchi\indices{^{\alpha'\beta'}_{\gamma'\delta'}}
  \dualF\indices{^{\gamma'\delta'}}
,
\end{align}
 where
~
\begin{align}
  \dualchi\indices{^{\alpha'\beta'}_{\gamma'\delta'}}
&=
  L\indices{^{\alpha'}_{\alpha}^{\beta'}_{\beta}}
 ~
  \dualchi\indices{^{\alpha\beta}_{\mu\nu}} 
 ~
  M\indices{^{\mu}_{\gamma'}^{\nu}_{\delta'}}
,\\
  L\indices{^{\alpha'}_{\alpha}^{\beta'}_{\beta}}
 &=
  {\Delta}^{-1}
  T\indices{^{\alpha'}_{\alpha}}
  T\indices{^{\beta'}_{\beta}}
,
\\
  M\indices{^{\gamma}_{\gamma'}^{\delta}_{\delta'}}
&=
  T\indices{^{\gamma}_{\gamma'}}
  T\indices{^{\delta}_{\delta'}}
.
\end{align}


Here we have seen how coordinate transformations can be
 actualized by seeing how they affect the constitutive tensor.
However,
 there is also another part to the wave mechanics --
 the differential equations.

If our coordinate transformation is not just a deformation 
 of our existing \emph{Cartesian} coordinate system, 
 but e.g. a change to cylindrical or polar coordinates, 
 then the wave equations change their form.
Although such radical coordinate conversions can be very useful, 
 they are a re-parameterization of the host space-time, 
 and not a straightforward deformation: 
 hence the appearance of extra factors of $r$ when converting
 from Cartesians to cylindrical coordinates.
For brevity, 
 we do not address this case here, 
 and focus instead on the deforming transformations
 used to produce particular T-devices.

%

Since the two field tensors might be compacted differently, 
 as happens for p-acoustics, 
 we need to allow for
 distinct compacted deformation matrices, 
 one for $\Tensor{\dualF}$, 
 and another for $\Tensor{G}$.
The compact deformation matrix equation
 for the vector $\MatrixB{\dualF^{B}}$
 depends on a compacted $M\indices{^{\alpha'}_{\alpha}^{\beta'}_{\beta}}$, 
 and is
~
\begin{align}
  \MatrixM{\dualF\indices{^{B'}}} 
&=
  \MatrixM{M\indices{^{B'}_{B}}}
  \MatrixM{\dualF\indices{^{B}}}
,
\end{align}
 but for the $\MatrixB{G\indices{^{A}}}$ we instead compact
 $L\indices{^{\alpha'}_{\alpha}^{\beta'}_{\beta}}$
 to get
~
\begin{align}
  \MatrixM{G\indices{^{A'}}}
&=
  \MatrixM{L\indices{^{A'}_{A}}}
  \MatrixM{G\indices{^{A}}}
.
\end{align}
The matrix representation
 of $\dualchi\indices{^{\alpha\beta}_{\gamma\delta}}$ 
 also transforms, 
 and is
~
\begin{align}
  \MatrixM{\dualchi'}
=
  \MatrixM{\dualchi\indices{^{A'}_{B'}}} 
&= 
  \MatrixM{L\indices{^{A'}_A}}
  \MatrixM{\dualchi\indices{^{A}_{B}}} 
  \MatrixM{M\indices{^{B'}_B}}^{-1}
.
\end{align}


So if we want to physically implement the transformation $T$
 as a T-device, 
 for any possible states of the field tensors $\Tensor{\dualF}$ and $\Tensor{G}$,
 we need to change the host material so that 
 its constitutive makeup is not simply the reference value $\dualchi$, 
 but that of the transformed constitutive makeup  $\dualchi'$.
In what follows we will implement this for both 
 purely spatial and the more exotic space-time cloaks
 in a ground-plane reference geometry.

In EM
 we would typically start (as we do below) 
 with a reference material and solution 
 based on a vacuum or a uniform static non-dispersive refractive index
 and -- for carpet T-devices --
 a light reflective surface.
For acoustics we would pick a stationary 
 and featureless elastic medium or fluid
 with an acoustically reflective surface.
After applying the deformation to the reference constitutive parameters, 
 we will have determined how the desired alterations
 to the wave mechanics (the field propagation)
 can be induced by appropriate changes to the propagation medium.

%
\section{Spatial Carpet T-devices}\label{S-carpet}

``Carpet'' or ground-plane transformation devices 
 are essentially surfaces that have been modified; 
 in their original conception they were engineered mirrors
 that appeared to give simple reflections but 
 were actually hiding some predefined region \cite{Li-P-2008prl}.
More generally,  
 though, 
 it is not required for the surface to be a mirror; 
 the transformation respecifies the properties inside 
 a region where the light propagates, 
 not what happens outside that region.
To make it clearer how our general prescription can be applied, 
 we will first apply it to the familiar spatial carpet cloak, 
 although generalized to incorporate other related T-devices.

\begin{figure}
\includegraphics[angle=-0,width=0.30\columnwidth]{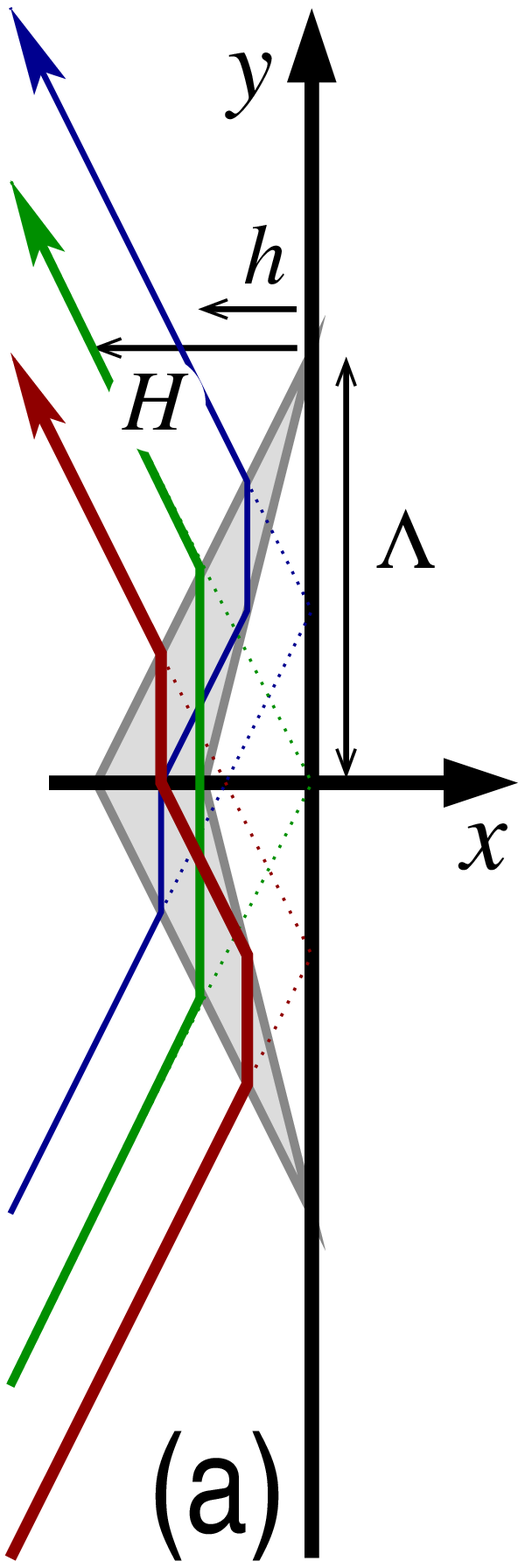}
\includegraphics[angle=-0,width=0.30\columnwidth]{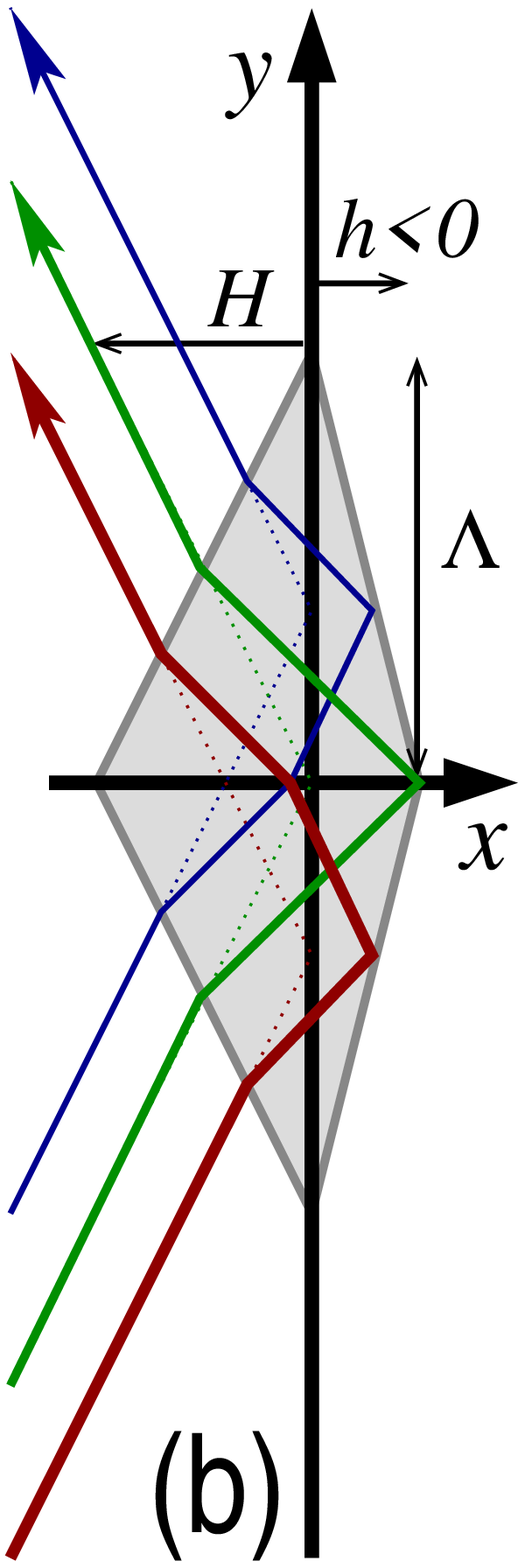}
\includegraphics[angle=-0,width=0.30\columnwidth]{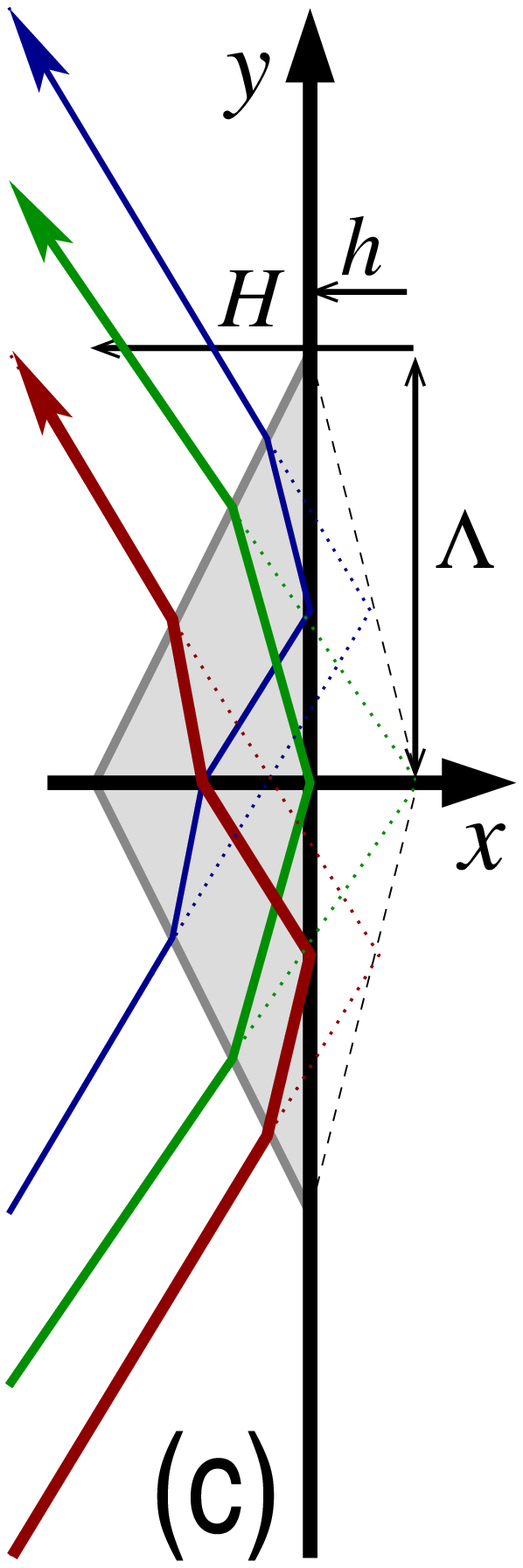}
\caption{
The (a) spatial carpet cloak, 
 (b) spatial {\peephole},
 (c) spatial tardis.
Continuous lines show the actual ray trajectories, 
 dotted lines the apparent (illusory) ray trajectories.
The {\peephole} (b) needs to be embedded in a high-index
 host medium to allow the internal rays to cover the 
 greater distance to the back surface
 than it is to the apparent surface.
}
\label{fig-carpet-Sray}
\end{figure}

We start with a planar carpet that lies flat on the $y-z$ plane 
 at $x=0$, 
 and decide to transform a localized region of space --
 the cloak ``halo'' -- 
 reaching no higher than $x=-H$ above the plane
 and no further than $y=\pm \Lambda$ sideways; 
 as shown by the shaded regions in Fig. \ref{fig-carpet-Sray}.
Firstly, 
 we can choose to restrict the space probed by incoming waves
 to that between the maximum height $H$ 
 and some lower height $x=h$, 
 but transform the material so that the \emph{apparent}
 space is expanded and extends all the way from $x=0$ to $x=-H$.
This T-device is the usual \textbf{carpet cloak}
 with an offset $h$ and scaling $R=(H-h)/H < 1$.
See Fig. \ref{fig-carpet-Sray}(a).

%
%

Secondly, 
 we might expand the actual space probed by incoming waves 
 to that between the maximum height $H$ 
 and to a penetration depth below the plane of  $x=h$, 
 but transform the material to shrink the \emph{apparent} space
 to only that between $x=0$ to $x=-H$.
This T-device is a cloaked ``\textbf{{\peephole}}''
 with a negative offset $h<0$ and scaling $R=(H-h)/H > 1$.
See Fig. \ref{fig-carpet-Sray}(b),
 where we see that the {\peephole} 
 allows an observer to remain \emph{below} the ground plane
 whilst allowing them an unrestricted view of their surroundings, 
 just as if they were exposed above it.


%
%

Thirdly, 
 we can define the actual space probed by incoming waves to be just the same
 as if there were a flat plane, 
 but nevertheless transform the material so that the \emph{apparent}
 space extends below the plane by $h$.
This T-device might be called a \textbf{tardis}, 
 since it presents the illusion that it 
 contains a space bigger than it is on the outside; 
 it uses an offset $h>0$ and scaling $R=(H-h)/H < 1$.  
Its diagram is shown in fig. \ref{fig-carpet-Sray}(c), 
 where the transformation is the reverse of that for the {\peephole}, 
 but in appearance it acts like a carpet cloak for when
 the observer expects to see a \emph{dimple} on the plane.


%
%
%
%

Lastly, 
 we could define the space probed by incoming waves to be just the same
 as if there were a flat plane, 
 but nevertheless transform the material so that the \emph{apparent}
 space was shrunk back as if there were a bump.
This T-device might be called an \textbf{anti-tardis},
 but for brevity we will not discuss such a device here.
Its diagram is not shown on fig. \ref{fig-carpet-Sray}, 
 but note that its transformation is the reverse of that for the carpet cloak 
 but in appearance it would act like a {\peephole} for when
 the observer expects to see a \emph{bump} on the plane.

The deformation that stretches a uniform medium 
 (un-primed coordinates) into 
 these T-device designs (primed coordinates)
 is applied (only) inside the shaded halo regions 
 on fig. \ref{fig-carpet-Sray},
 and deforms only $x$ so that 
~
\begin{align}
  t' &= t,
\\
  x' &= R x - C \frac{\Lambda - y.\sgn(y)}{\Lambda} h 
  \nonumber
\\
     &= R x - C h + r s y,
\label{eqn-scarpet-deform}
\\
y' &= y,
\\
z' &= z
,
\end{align} 
 where the ratio of actual depth ($H-h$)
 to apparent depth ($H$)
 is $R=(H-h)/H$, 
 and $r=Ch/\Lambda$,
 and $s=\sgn(y)$.
Here $C$ is set to one
 for the carpet cloak and {\peephole}, 
 but for the tardis (where $R>1$)
 it needs to be $C=R$.
Using a general viewpoint applicable to any of these three T-devices,
 we see that 
 $R<1$ apparently expands the wave-accessible space, 
 as needed for a cloak or tardis, 
 whilst $R>1$ compresses it, 
 as needed for a {\peephole} or anti-tardis.

The effect of this deformation from eqn. \eqref{eqn-scarpet-deform}
 is given
 by $T\indices{^{\alpha '}_{\alpha}}$, 
 which specifies the differential relationships between the 
 primed and un-primed coordinates, 
 where $\alpha \in \{ t,x,y,z \}$ and $\alpha' \in \{ t',x',y',z' \}$.
If we let rows span the first (upper) index, 
 and columns the second (lower) index,
 then
~
\begin{align}
  T\indices{^{\alpha '}_{\alpha}}
&=
  \MatrixM{
    \frac{\partial x^{\alpha'}}{\partial x^\alpha}
  }
&=&
    \begin{bmatrix}
      1  &  0  &  0  &  0  \\
     ~0~ & ~R~ &  rs & ~0~ \\
      0  &  0  &  1  &  0  \\
      0  &  0  &  0  &  1
    \end{bmatrix}
,
\end{align}
and note that $\det (T\indices{^{\alpha '}_{\alpha}}) = R$.
For some column 4-vector $V^\alpha$, 
 we have that 
~
\begin{align}
  V\indices{^{\alpha '}}
&=
  T\indices{^{\alpha '}_{\alpha}}
  V\indices{^{\alpha}},
\\
\textrm{i.e.}\qquad
    \begin{bmatrix}
      V\indices{^{t'}} \\
      V\indices{^{x'}} \\
      V\indices{^{y'}} \\
      V\indices{^{z'}}
    \end{bmatrix}
&=
    \begin{bmatrix}
      1  &  0  &  0  &  0  \\
     ~0~ & ~R~ &  rs & ~0~ \\
      0  &  0  &  1  &  0  \\
      0  &  0  &  0  &  1
    \end{bmatrix}
    \begin{bmatrix}
      V\indices{^t} \\
      V\indices{^x} \\
      V\indices{^y} \\
      V\indices{^z}
    \end{bmatrix}
.
\end{align}

It is worth noting that inside the carpet cloak halo -- 
 where the cloak affects the path of the rays, 
 but not inside the cloak core where objects are hidden, 
 the deformation consists solely of a constant rescaling of $x$; 
 the rest of the cloak design is defined by its spatial layout, 
 reversed properties in the upper and lower halves, 
 and its boundaries.

%
\subsection{EM carpets}\label{SS-carpet-EM}

We expressed the EM constitutive tensor 
 using a compact $6 \times 6$ matrix form,
 therefore we also use compact $6 \times 6$ 
 deformation (transformation) matrices $\MatrixB{L}, \MatrixB{M}$, 
 being versions of the product $L\indices{^{\alpha '}_{\alpha}}
  L\indices{^{\beta '}_{\beta}}$.
Note that the use of the unconventional choice 
 of (the dual) $\dualF\indices{^{\alpha\beta}}$
 and (the mixed) $\dualchi\indices{^{\alpha\beta}_{\gamma\delta}}$
 allows us to calculate using straightforward matrix multiplication.

The 
 two dimensional matrix representation
 of the transformation matrix, 
 as compressed for EM,
 is best written as an explicit transformation between 
 the field 6-vectors to avoid error.
Thus the compressed column 6-vector $\MatrixB{G\indices{^{A}}}$
 is transformed using\footnote{Remember that for EM, 
  each element of $\MatrixB{L}$ 
  not only includes the obvious 
  contributions to $G\indices{^{{\alpha'}{\beta'}}}$
 from $G\indices{^{\alpha\beta}}$, 
  but also has a sign-flipped one from $G\indices{^{\beta\alpha}}$.
 This accounts for the $-rs$ entry in $\MatrixB{L}$.}
~
\begin{align}
  \MatrixM{ G\indices{^{A'}} }
&=
  \MatrixM{ L\indices{^{A'}_{A}} }
  \MatrixM{ G\indices{^{A}} }
\\
    \begin{bmatrix}
      G\indices{^{t'x'}} \\
      G\indices{^{t'y'}} \\
      G\indices{^{t'z'}} \\
      G\indices{^{y'z'}} \\
      G\indices{^{z'x'}} \\
      G\indices{^{x'y'}} 
    \end{bmatrix}
&=
  \frac{1}{R}
    \begin{bmatrix}
      R  & +rs &  0  &  0  &  0  &  0  \\
      0  &  1  &  0  &  0  &  0  &  0  \\
      0  &  0  &  1  &  0  &  0  &  0  \\
      0  &  0  &  0  &  1  &  0  &  0  \\
      0  &  0  &  0  & -rs &  R  &  0  \\
      0  &  0  &  0  &  0  &  0  &  R  
    \end{bmatrix}
    \begin{bmatrix}
      G\indices{^{tx}} \\
      G\indices{^{ty}} \\
      G\indices{^{tz}} \\
      G\indices{^{yz}} \\
      G\indices{^{zx}} \\
      G\indices{^{xy}} 
    \end{bmatrix}
,
\end{align}
 where the matrix $\MatrixB{M}$ that transforms $\MatrixB{\dualF^{B}}$ 
 is the same.
To transform $\MatrixB{\dualchi}$, 
 we need the inverse of $\MatrixB{M}$, 
 which is 
~
\begin{align}
  \MatrixM{ M\indices{^{A'}_{A}} }^{-1}
&=
    \begin{bmatrix}
     ~1~ & -rs & ~0~ & ~0~ & ~0~ & ~0~ \\
      0  &  R  &  0  &  0  &  0  &  0  \\
      0  &  0  &  R  &  0  &  0  &  0  \\
      0  &  0  &  0  &  R  &  0  &  0  \\
      0  &  0  &  0  & +rs &  1  &  0  \\
      0  &  0  &  0  &  0  &  0  &  1
    \end{bmatrix}
.
\end{align}

Starting from simple uniform background medium
 described only by $\epsilon$ and $\mu$ --
 perhaps a vacuum with $\epsilon=\epsilon_0$ and $\mu=\mu_0$ --
 the transformed constitutive matrix
 defining the EM carpet T-device is then
~
\begin{align}
&
  \MatrixM{\dualchi\indices{^{A'}_{B'}}} 
 = 
    \MatrixM{ L\indices{^{A'}_{A}} }
    \MatrixM{ \dualchi\indices{^{A}_{B}} }
    \MatrixM{ M\indices{^{B'}_{B}} }^{-1}
\\
&=
  \frac{\epsilon}{R}
    \begin{bmatrix} 
      ~0~    &  ~0~         &  ~0~    &-(R^2+r^2) & -rs   &  0 \\
      ~0~    &  ~0~         &   0     & -rs       &  1    &  0 \\
      ~0~    &   0          &  ~0~    &  0        &  0    &  1 \\
     ~c^2    & -rs c^2      &  ~0~    &  0        &  0    &  0 \\
     -rs c^2 & c^2 (R^2+r^2)&  ~0~    &  0        &  0    &  0 \\
      ~0~    &  ~0~         & R^2 c^2 &  0        &  0    &  0 \\
    \end{bmatrix}
.
\end{align}

\begin{figure}
\includegraphics[angle=-0,width=0.32\columnwidth]{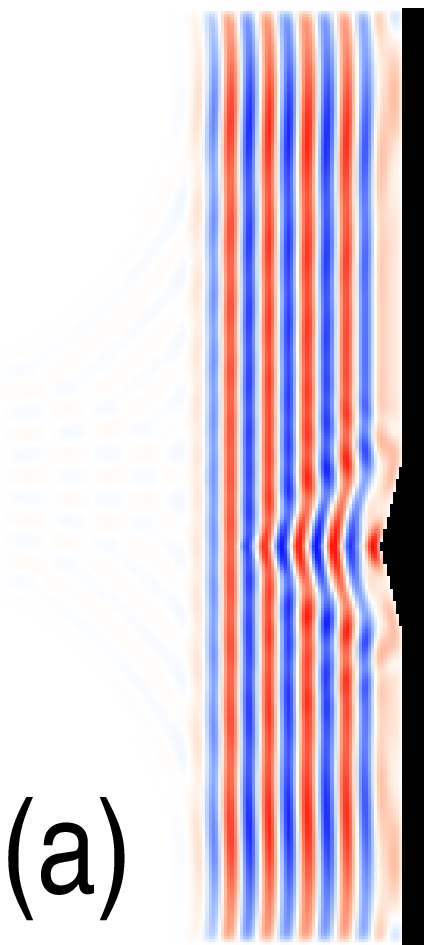}
\includegraphics[angle=-0,width=0.32\columnwidth]{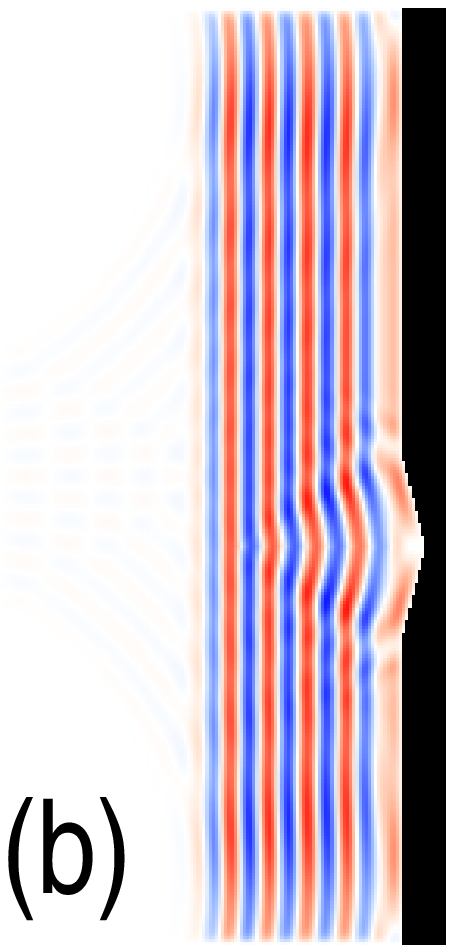}
\includegraphics[angle=-0,width=0.32\columnwidth]{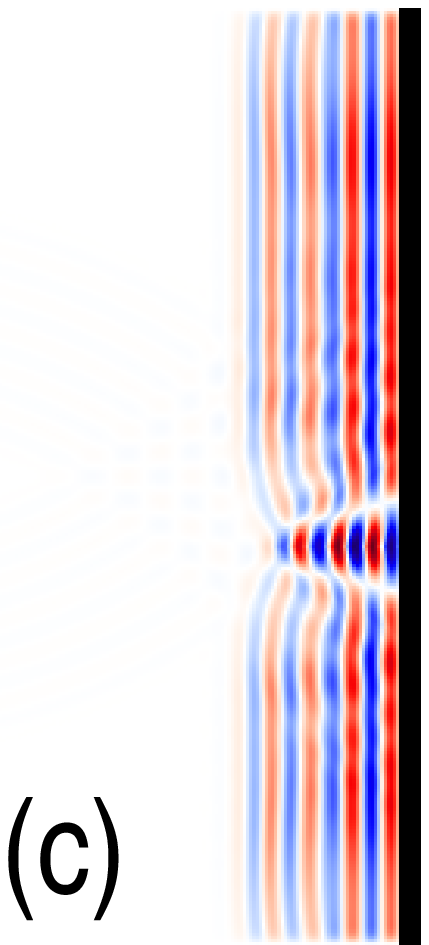}
\caption{
EM finite difference time domain (FDTD) snapshots of 
 the (a) spatial carpet cloak, 
 (b) spatial {\peephole},
 (c) spatial tardis.
The cloak and {\peephole} frames (a,b) 
 show a burst of exactly perpendicular plane waves
 just as the leading edge hits the 
 mirror surface.
This makes it clear how the cloaking transformation 
 has ensured the wavefront conforms perfectly to the mirror profile.
In (c), 
 we see the burst of plane waves \emph{after} reflection, 
 where the extra virtual space created by the tardis 
 transformation has imprinted a dent on the wavefront.
}
\label{fig-carpet-Sray-FDTD}
\end{figure}

So we see a need for birefringence, 
 as defined by the off-diagonal permittivity elements $-\epsilon rs/R$, 
 and the off-diagonal permeability contributions $-\epsilon rs c^2/R$.
To get a $\mu$ matrix, 
 we need to invert the $\eta$ sub-matrix:
~
\begin{align}
[\mu]
=
    \begin{bmatrix}
     1/R    &  -rs/R   &  ~0~  \\
     -rs/R  &  1+r^2/R &  ~0~  \\
     ~0~    &  ~0~     &   R \\
    \end{bmatrix}^{-1}
&=
    \frac{1}{R}
    \begin{bmatrix}
       R^2+r^2 &  +rs  &  ~0~  \\
      +rs      &  ~1~  &  ~0~  \\
      ~0~      &  ~0~  &   1  \\
    \end{bmatrix}
.
\end{align}
Note how the structure of the (sub)matrix $\MatrixB{\mu}$ 
 matches up with that of the (sub)matrix $\MatrixB{\epsilon}$,
 as we would expect as our transformation process 
 demands a preservation of impedance matching.
Numerical examples indicating how the EM fields are 
 distorted by the different carpet T-devices 
 are shown on fig. \ref{fig-carpet-Sray-FDTD}.

As is well known, 
 the requirement for impedance matching can be relaxed.
Since the simple nature of the transformation
 of eqn. \eqref{eqn-scarpet-deform}
 leads to a constitutive matrix with constant values, 
 this means that these carpet devices can be made 
 for a specified polarization 
 using the natural birefringence
 of two correctly shaped and oriented pieces of calcite -- 
 as already demonstrated for a carpet cloak 
 \cite{Chen-LZJPZ-2011nc,Zhang-LLB-2011prl}.



%
\subsection{p-Acoustics carpets}\label{SS-carpet-pA}

For p-acoustic transformations, 
 we need two compactified transformation matrices, 
 with correctly arranged ordering.
This is less straightforward than in EM, 
 where rows and columns are indexed by the same list of coordinate pairs.
The vector $\MatrixB{G\indices{^A}}$ 
 is most compactly indexed by $ww,xt,yt,zt$, 
 where ``$w$'' stands in for one of $x, y$ or $z$.
Where this means that some elements may have multiple values, 
 we will indicate this using (e.g.) $\{R,1,1\}$
 for $x$, $y$, and $z$ choices respectively.
To transform this we need $\MatrixB{L\indices{^{A'}_A}}$, 
 which is\footnote{Since $w$ represents multiple coordinate choices,
 the $w'w'=x'x'$ element of $\MatrixB{L}$, 
 i.e. $L\indices{^{w'w'}_{ww}} = R^2+r^2$, 
  results from a sum over all possible $ww \in \{xx,yy,zz\}$,
  in combination with $G\indices{^{xx}}=G\indices{^{yy}}=p$.}
~
\begin{align}
  \MatrixM{ G\indices{^{A'}} }
&=
  \MatrixM{ L\indices{^{A'}_{A}} }
  \MatrixM{ G\indices{^{A}} }
\\
    \begin{bmatrix}
      G\indices{^{w'w'}} \\
      G\indices{^{x't'}} \\
      G\indices{^{y't'}} \\
      G\indices{^{z't'}}
    \end{bmatrix}
&=
  \frac{1}{R}
    \begin{bmatrix}
    \{R^2+r^2,1,1\}  &   0  &  0  &  0  \\
     0   &   R  &  rs &  0  \\
     0   &   0  &  1  &  0  \\
     0   &   0  &  0  &  1  \\
    \end{bmatrix}
    \begin{bmatrix}
      G\indices{^{ww}} \\
      G\indices{^{xt}} \\
      G\indices{^{yt}} \\
      G\indices{^{zt}}
    \end{bmatrix}
.
\end{align}

To transform $\MatrixB{\dualF\indices{^B}}$,
 compactly indexed by $tt, xt, yt, zt$,
 we need $\MatrixB{M\indices{^{B'}_B}}$ 
 (and also its inverse),
~
\begin{align}
  \MatrixM{ \dualF\indices{^{B'}} }
&=
  \MatrixM{ M\indices{^{B'}_{B}} }
  \MatrixM{ \dualF\indices{^{B}} }
\\
    \begin{bmatrix}
      F\indices{^{t't'}} \\
      F\indices{^{x't'}} \\
      F\indices{^{y't'}} \\
      F\indices{^{z't'}}
    \end{bmatrix}
&=
  \frac{1}{R}
    \begin{bmatrix}
     1  &   0  &  0  &  0  \\
     0  &   R  &  rs &  0  \\
     0  &   0  &  1  &  0  \\
     0  &   0  &  0  &  1  \\
    \end{bmatrix}
    \begin{bmatrix}
      F\indices{^{tt}} \\
      F\indices{^{xt}} \\
      F\indices{^{yt}} \\
      F\indices{^{zt}}
    \end{bmatrix}
,
\end{align}

Starting from the simplest acoustic medium, 
 i.e. one for which $\Vec{\alpha}=\Vec{\beta}=\Vec{0}$ 
  and $\Mat{\rho} \rightarrow \rho$,
 the transformed constitutive matrix
 defining the p-acoustic carpet T-device is then
~
\begin{align}
  \MatrixM{\dualchi\indices{^{A'}_{B'}}}
&=
    \MatrixM{L\indices{^{A'}_{A}}}
    \MatrixM{\dualchi\indices{^{A}_{B}}}
    \MatrixM{M\indices{^{B'}_{B}}}^{-1}
\\
&=
  \rho
    \begin{bmatrix}
   \{(R^2+r^2),1,1\}c^2 & ~~0~~ & ~~0~~ & ~~0~~ \\
      0                 &   1   &   0   &   0  \\
      0                 &   0   &   1   &   0  \\
      0                 &   0   &   0   &   1
    \end{bmatrix}
,
\end{align}
 since $c^2 = \kappa/\rho$.

Here we can immediately see that there is no way
 of constructing a general carpet cloak
 (or related T-device) with p-acoustic waves.
This is because the
 transformed energy parameter  $\kappa'$
 is required to have mutually incompatible values:
 i.e. both $(R^2+r^2)\rho c^2$ and $\rho c^2$ at the same time!
We can avoid this incompatibility
 by restricting ourselves to only 1D $x$-axis propagation, 
 so that the waves need experience only one of these $\kappa'$ values.
Although such a 1D implementation might technically match
 the original conception of a carpet cloak, 
 it is merely a static waveguide that appears longer than it really is.
However, 
 as we will see below, 
 a linear \emph{space-time} carpet T-device can be more interesting.


%
\section{Space-time carpet T-devices}\label{S-stcarpet}

The basic design used here is geometrically related 
 to that described above, 
 except that one of the two spatial coordinates
 is replaced by the time coordinate.
This means that waves are not diverted around the event
 in a spatial sense, 
 instead the leading waves 
 (in an optical cloak, 
  the ``illumination'')
 are sped up, 
 and the latter slowed down, 
 creating a ``dark'' wave-free interval in which timed events are obscured
 \cite{McCall-FKB-2011jo,Kinsler-M-2014adp-scast}.
After the event, 
 the leading part of the illumination (whether sound or light) is slowed, 
 and the latter sped up until the initial uniform, 
 seamless illumination has been perfectly reconstructed.
For an acoustics implementation, 
 we should think of sonar-using creatures such as bats or dolphins
 who rely on incoming sound waves to locate and observe events; 
 an acoustic event cloak creates a quiet region
 from which no sound reflections can be heard,
 all without leaving any tell-tale distortion of
 either the background soundscape or deliberate sonar ``pings''.

\begin{figure}
\includegraphics[angle=-0,width=0.32\columnwidth]{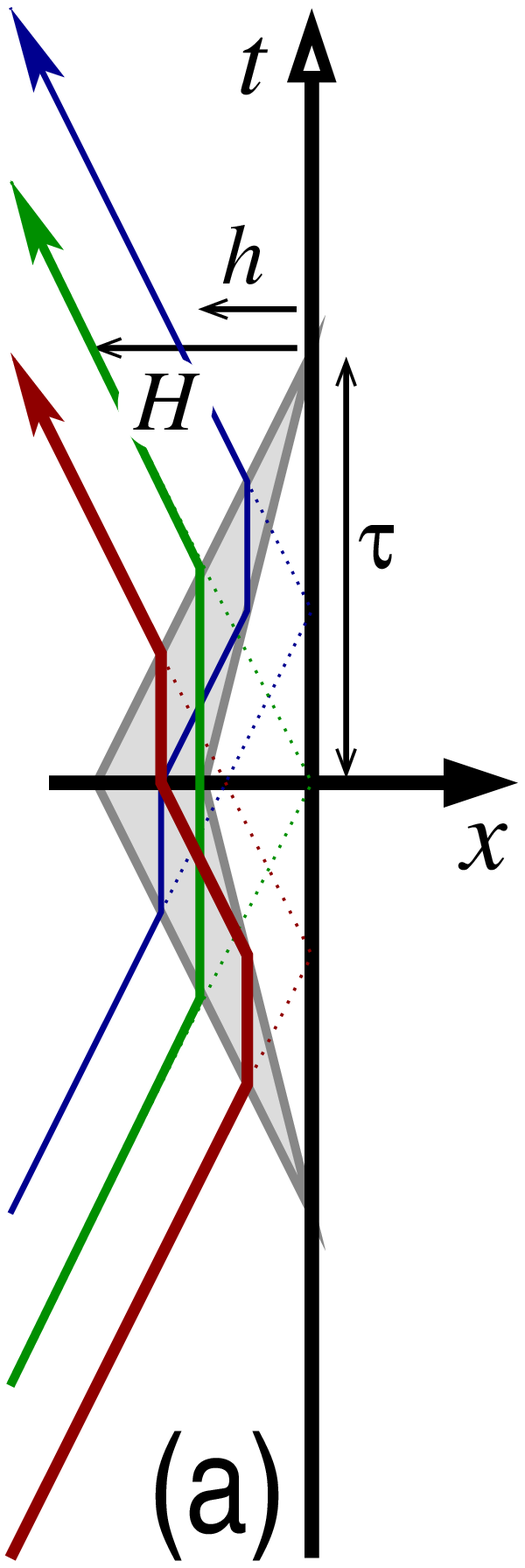}
\includegraphics[angle=-0,width=0.32\columnwidth]{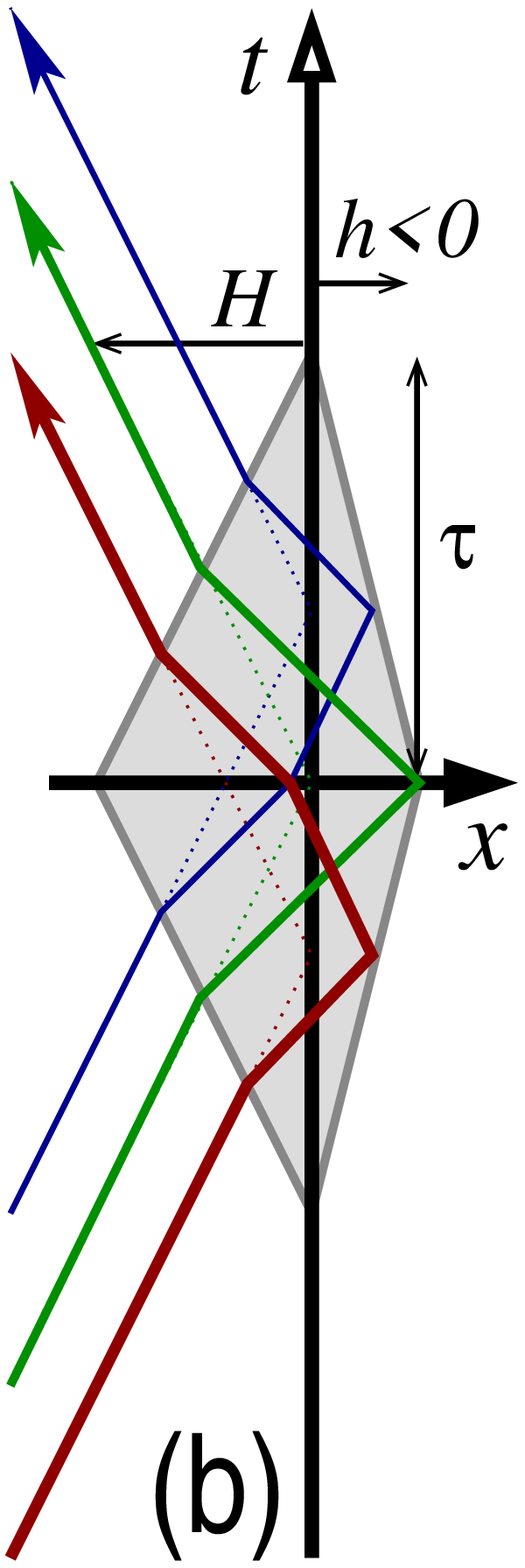}
\includegraphics[angle=-0,width=0.32\columnwidth]{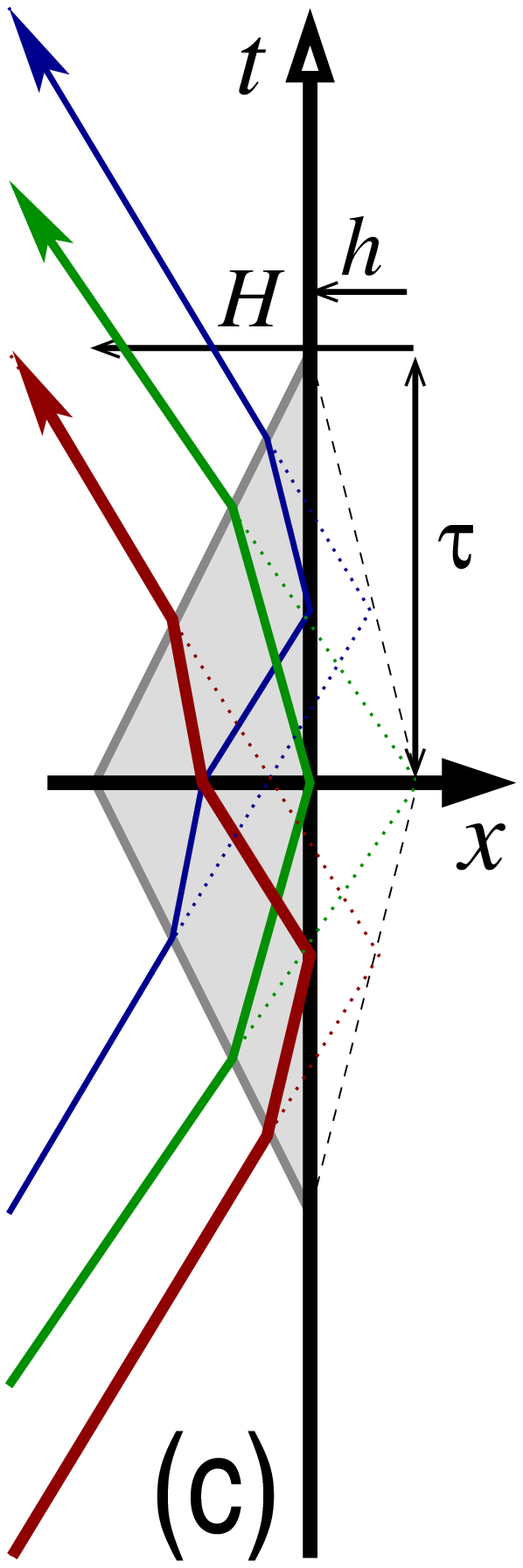}
\caption{
The (a) space-time ``event'' carpet cloak, 
 (b) space-time {\peephole},
 (c) space-time tardis.
Continuous lines show the actual ray trajectories, 
 dotted lines the apparent (illusory) ray trajectories.
The {\peephole} (b) needs to be embedded in a high-index
 host medium to allow the internal rays to cover the 
 greater distance to the back surface
 than it is to the apparent surface.
For the carpet cloak diagram (a), 
 we have exaggerated the transformation so that
 some rays are temporarily stopped -- 
 but note that any realistic implementation
 will involve much weaker speed modulations.
}
\label{fig-carpet-Tray}
\end{figure}

We assume a carpet that lies flat on the $y-z$ plane 
 at $x=0$, 
 and decide to cloak a region $h$ deep and $\tau$ in duration, 
 with the cloak's influence extending out as far as $H+h$ above the carpet.
The deformation that stretches a uniform and static medium 
 (un-primed coordinates) into 
 these T-device designs (primed coordinates)
 is applied (only) inside the shaded ``halo'' regions 
 on fig. \ref{fig-carpet-Tray}, 
 and deforms only $x$ so that 
~
\begin{align}
  t' &= t,
\\
  x' &= \frac{H-h}{H} x - C \frac{\tau - t.\sgn(t)}{\tau} h
  \nonumber
\\
     &= R x - C h + r s t,
\label{eqn-tcarpet-deform}
\\
y' &= y,
\\
z' &= z
,
\end{align} 
 where $R=(H-h)/H$ is the ratio of actual depth to apparent depth, 
 and now $r=Ch/\tau$, 
 and $s=\sgn(t)$. 
Here C is usually set to one; 
 except for the space-time tardis, 
 where it is $R$.

As in the spatial case, 
 as well as the cloak shown on Fig. \ref{fig-carpet-Tray}(a),
 this deformation can also represent other T-devices, 
 such as
 a \emph{space-time {\peephole}}, 
  as shown on Fig. \ref{fig-carpet-Tray}(b),
 and a \emph{space-time tardis}, 
  as shown on Fig. \ref{fig-carpet-Tray}(c), 
 or perhaps even a {space-time anti-tardis}.
The space-time {\peephole} 
 allows an observer to remain always \emph{below} the ground plane
 whilst temporarily allowing them an unrestricted view of their surroundings, 
 just as if they briefly put their head out above it; 
 it is specified using $h<0$.
Thus, 
 for example,
 a sensor can be briefly exposed to take readings
 whilst usually remaining in shelter below the plane.
The tardis gives the onlooker the temporary impression of a space
 bigger than it actually is,
 with the ground plane appearing to temporarily recede,
 this requires that $C=R$.
Although in this carpet implementation the space-time tardis 
 is the same as a space-time carpet cloak for a dimple, 
 in a non-carpet radial implementation
 the visual effect is decidedly more startling!

Using a general viewpoint applicable to any of these three T-devices,
 we see that $H$ is the apparent depth detectable by an observer, 
 positive-valued $h$ apparently expands the accessible space, 
 and negative $h$ apparently shrinks it.

To ensure that the deformed solution includes no
 waves travelling faster than the maximum wave speed $c$,
 we also specify that the aspect ratio $(H-h)/c\tau$
 is less than one.
As seen on fig. \ref{fig-carpet-Tray}(a),
 this deformation pushes a small triangular space-time region --
 the hidden cloaked region --
 of the wave illumination away from the carpet plane,
However, 
 note that if \emph{only} the spatial extent is considered, 
 the entire ground plane itself moves out and then back
 to temporarily hide a slab of space.

Despite all this trickery, 
 as the (incoming) waves approach the carpet plane from the (below) left, 
 any deviations from ordinary straight-line propagation 
 that occur within the shaded region 
 are compensated for, 
 so that when (outgoing) waves emerge travelling towards the (top) left,
 they appear to have reflected from a flat carpet plane.
Note that the cloaking device reverses the speed changes
 at $t=0$:
 on the figure, 
 at early times $t<0$, 
 incoming waves are slowed or stopped and outgoing waves are unaffected,
 but at late times $t>0$, 
 it is the outgoing waves that are slowed or stopped, 
 with incoming waves travelling normally.
For the practical situation, 
 i.e. cloaks where $(H-h)>h$, 
 waves are not stopped but slowed.

The effect of this deformation is given
 by the transformation matrix 
 $T\indices{^{\alpha '}_{\alpha}}$, 
 which specifies the differential relationships
 between the deformed (primed)
 and reference (unprimed) coordinates, 
 where $\alpha \in \{ t,x,y,z \}$ and $\alpha' \in \{ t',x',y',z' \}$.
If we let rows span the first (upper) index, 
 and columns the second (lower) index,
 then
~
\begin{align}
  T\indices{^{\alpha '}_{\alpha}}
&=
  \MatrixM{
    \frac{\partial x^{\alpha'}}{\partial x^\alpha}
  }
\quad =
    \begin{bmatrix}
      1  &  0  &  0  &  0  \\
      rs &  R  &  0  &  0  \\
      0  &  0  &  1  &  0  \\
      0  &  0  &  0  &  1
    \end{bmatrix}
,
\end{align}
and note that $\det (T\indices{^{\alpha '}_{\alpha}}) = R$.
For some column 4-vector $V^\alpha$, 
 we have that 
~
\begin{align}
  V\indices{^{\alpha '}}
&=
  T\indices{^{\alpha '}_{\alpha}}
  V\indices{^{\alpha}}
,
\\
 \textrm{i.e.} \qquad
    \begin{bmatrix}
      V\indices{^{t'}} \\
      V\indices{^{x'}} \\
      V\indices{^{y'}} \\
      V\indices{^{z'}}
    \end{bmatrix}
&=
    \begin{bmatrix}
      1  &  0  &  0  &  0  \\
      rs &  R  &  0  &  0  \\
      0  &  0  &  1  &  0  \\
      0  &  0  &  0  &  1
    \end{bmatrix}
    \begin{bmatrix}
      V\indices{^t} \\
      V\indices{^x} \\
      V\indices{^y} \\
      V\indices{^z}
    \end{bmatrix}
.
\end{align}

Just as for the spatial carpet T-devices
 in section \ref{S-carpet}, 
 the deformation consists solely of a constant rescaling of $x$
 and so the transformed constitutive parameters again have fixed values.
However, 
 in contrast to the spatial case,
 in these space-time T-devices 
 the constitutive parameters toggle between fixed values,
 and boundaries \emph{move}.
This will be discussed in section \ref{S-implementation}.

%
\subsection{EM event carpets}\label{SS-stcarpet-EM}

We expressed the EM constitutive tensor 
 using a compact $6 \times 6$ matrix form,
 therefore we also use compact $6 \times 6$ 
 deformation (transformation) matrices $\MatrixB{L}, \MatrixB{M}$, 
 being versions of the product $L\indices{^{\alpha '}_{\alpha}}
 L\indices{^{\beta '}_{\beta}}$.
Note that the use of the unconventional choice 
 of (the dual) $\dualF\indices{^{\alpha\beta}}$
 and (the mixed) $\dualchi\indices{^{\alpha\beta}_{\gamma\delta}}$
 allows us to calculate using straightforward matrix multiplication.

The two dimensional matrix representation of the transformation matrix,
 as compressed for EM,
 is best written as an explicit transformation between the field 6-vectors
 to avoid error. 
The compressed column 6-vector $\MatrixB{G\indices{^A}}$
 is transformed using\footnote{Remember that for EM,
  each element of $\MatrixB{L}$ 
  not only includes the obvious 
  contributions to
  $G\indices{^{{\alpha'}{\beta'}}}$ from $G\indices{^{\alpha\beta}}$, 
  but also has a sign-flipped one from $G\indices{^{\beta\alpha}}$.}
~
\begin{align}
  \MatrixM{G\indices{^{A'}}}
&=
  \MatrixM{L\indices{^{A'}_{A}}}
  \MatrixM{G\indices{^{A}}}
\\
    \begin{bmatrix}
      G\indices{^{t'x'}} \\
      G\indices{^{t'y'}} \\
      G\indices{^{t'z'}} \\
      G\indices{^{y'z'}} \\
      G\indices{^{z'x'}} \\
      G\indices{^{x'y'}} 
    \end{bmatrix}
&=
  \frac{1}{R}
    \begin{bmatrix}
     ~R~ & ~0~ & ~0~ & ~0~ & ~0~ & ~0~ \\
      0  &  1  &  0  &  0  &  0  &  0  \\
      0  &  0  &  1  &  0  &  0  &  0  \\
      0  &  0  &  0  &  1  &  0  &  0  \\
      0  &  0  & -rs &  0  &  R  &  0  \\
      0  & +rs &  0  &  0  &  0  &  R  
    \end{bmatrix}
    \begin{bmatrix}
      G\indices{^{tx}} \\
      G\indices{^{ty}} \\
      G\indices{^{tz}} \\
      G\indices{^{yz}} \\
      G\indices{^{zx}} \\
      G\indices{^{xy}} 
    \end{bmatrix}
,
\end{align}
 and
 where the matrix $\MatrixB{M}$ that transforms $\MatrixB{\dualF^{B}}$
 is the same. 
To transform $\MatrixB{\dualchi}$, 
 we need the inverse of $\MatrixB{M}$, 
 which is 
~
\begin{align}
  \MatrixM{M\indices{^{B'}_{B}}}^{-1}
&=
    \begin{bmatrix}
     ~1~ & ~0~ & ~0~ & ~0~ & ~0~ & ~0~ \\
      0  &  R  &  0  &  0  &  0  &  0  \\
      0  &  0  &  R  &  0  &  0  &  0  \\
      0  &  0  &  0  &  R  &  0  &  0  \\
      0  &  0  & +rs &  0  &  1  &  0  \\
      0  & -rs &  0  &  0  &  0  &  1  
    \end{bmatrix}
.
\end{align}

Starting from simple background medium
 described only by scalar $\epsilon$ and $\mu$, 
 the transformed constitutive matrix
 defining the EM space-time carpet T-device is then
~
\begin{align}
&  \MatrixM{\dualchi\indices{^{A'}_{B'}}}
=
    \MatrixM{ L\indices{^{A'}_{A}} }
    \MatrixM{ \dualchi\indices{^{A}_{B}}}
    \MatrixM{ M\indices{^{B'}_{B}} }^{-1}
\\
&=
  \frac{\epsilon}{R}
    \begin{bmatrix}
      ~0~  &  ~0~       &  ~0~       & -R^2       &  0          &  0      \\
      ~0~  &  ~0~       &  -rs       &  0         & -1          &  0      \\
      ~0~  &  +rs       &  ~0~       &  0         &  0          & -1      \\
      c^2  &  ~0~       &  ~0~       &  0         &  0          &  0      \\
      ~0~  & R^2c^2-r^2 &  ~0~       &  0         &  0          & +rs     \\
      ~0~  &  ~0~       & R^2c^2-r^2 &  0         & -rs         &  0      \\
    \end{bmatrix}
.
\end{align}
Here we see the slowing/speeding behaviour
 of an ordinary space-time cloak \cite{McCall-FKB-2011jo}, 
 but with a direction-sensitivity 
 caused by magneto-electric effects $\pm rs \epsilon /R$ in the material.
This direction-sensitivity reverses at $t=0$,
 so that while the incoming light travels slower than outgoing light
 when $t<0$, 
 the incoming light travels faster for $t>0$.
An implementation of this design, 
 will be discussed in section \ref{S-implementation}.


%
\subsection{p-Acoustics event carpets}\label{SS-stcarpet-pA} 

For p-acoustic transformations, 
 we need two compactified transformation matrices, 
 with correctly arranged ordering.
Again,
 and unlike EM,
 the compacted 
 $\MatrixB{G\indices{^A}}$ is indexed by $ww,xt,yt,zt$, 
 where ``$w$'' stands in for one of $x, y$ or $z$; 
 and any multiple valued elements are indicated using 
 a set-like notation. 
We need $\MatrixB{L\indices{^{A'}_A}}$
 to transform $\MatrixB{G\indices{^A}}$, 
 and it is
~
\begin{align}
  \MatrixM{ G\indices{^{A'}} }
&=
  \MatrixM{ L\indices{^{A'}_{A}} }
  \MatrixM{ G\indices{^{A}} }
\\
    \begin{bmatrix}
      G\indices{^{w'w'}} \\
      G\indices{^{x't'}} \\
      G\indices{^{y't'}} \\
      G\indices{^{z't'}}
    \end{bmatrix}
&=
  \frac{1}{R}
    \begin{bmatrix}
    \{R^2,1,1\} & \{rsR,0,0\} &  0  &  0  \\
      0  &   R  &  0  &  0  \\
      0  &   0  &  1  &  0  \\
      0  &   0  &  0  &  1  \\
    \end{bmatrix}
    \begin{bmatrix}
      G\indices{^{ww}} \\
      G\indices{^{xt}} \\
      G\indices{^{yt}} \\
      G\indices{^{zt}}
    \end{bmatrix}
.
\end{align}

The compact $\MatrixB{\dualF\indices{^A}}$,
 indexed by $tt, xt, yt, zt$,
 is transformed using 
 $\MatrixB{M\indices{^{B'}_B}}$ 
~
\begin{align}
  \MatrixM{ \dualF\indices{^{B'}} }
&=
  \MatrixM{ M\indices{^{B'}_{B}} }
  \MatrixM{ \dualF\indices{^{B}} }
\\
    \begin{bmatrix}
      F\indices{^{t't'}} \\
      F\indices{^{x't'}} \\
      F\indices{^{y't'}} \\
      F\indices{^{z't'}}
    \end{bmatrix}
&=
  \frac{1}{R}
    \begin{bmatrix}
      1  &   0  &  0  &  0  \\
      rs &   R  &  0  &  0  \\
      0  &   0  &  1  &  0  \\
      0  &   0  &  0  &  1  \\
    \end{bmatrix}
    \begin{bmatrix}
      F\indices{^{tt}} \\
      F\indices{^{xt}} \\
      F\indices{^{yt}} \\
      F\indices{^{zt}}
    \end{bmatrix}
.
\end{align}
where
~
\begin{align}
  \MatrixM{ M\indices{^{B'}_{B}} }^{-1}
&=
  \MatrixM{ M\indices{^{B}_{B'}} }
&=
    \begin{bmatrix}
      R  &   0  &  0  &  0  \\
     -rs &   1  &  0  &  0  \\
      0  &   0  &  R  &  0  \\
      0  &   0  &  0  &  R  \\
    \end{bmatrix}
.
\end{align}

Hence, 
 since $c^2=\kappa/\rho$, 
 and as usual starting with a simple background medium
 described only by $\kappa$ and $\rho$, 
 the transformed constitutive matrix $\MatrixB{\dualchi'}$
 defining the p-acoustic space-time carpet T-device is 
~
\begin{align}
  \MatrixM{L\indices{^{A'}_{A}}}
  \MatrixM{\dualchi\indices{^{A}_{B}}}
  \MatrixM{M\indices{^{B'}_{B}}}^{-1}
&=
  \rho
    \begin{bmatrix}
    \{R^2-\frac{r^2}{c^{2}},1,1\}c^2 & \{rs,0,0\} & ~0~          & ~0~ \\
     -rs        &   1            &  0           &  0  \\
      0         &   0            &  1           &  0  \\
      0         &   0            &  0           &  1  \\
    \end{bmatrix}
.
\end{align}

Again we can see that there is no way of constructing
 a general space-time carpet cloak (or related T-device)
 with p-acoustic waves,
 which we can see from the fact that 
 the modulus $\kappa'$ and $\alpha'_x$ cross-coupling
 are both required to have two incompatible values. 
In contrast to the spatial carpet cloak, 
 a one dimensional (in $x$) space-time carpet cloak
 can make sense; 
 just as the one dimensional space-time EM cloak
 makes sense \cite{McCall-FKB-2011jo,Fridman-FOG-2012nat,Lukens-LW-2013n,Lerma-2012preprint}.
However, 
 the necessarily bi-directional nature of the design
 means that we would require 
 a controllable coupling $\alpha'_x=rs\rho$ 
  enabling the velocity field $v_x \equiv \dualF^{tx}$
  to affect the pressure $p \equiv G^{xx}$, 
 as well as a $\beta'_x=-rs\rho$ 
  enabling the density $P \equiv \dualF^{tt}$
  to affect the momentum density $\Co{v}_x \equiv G^{xt}$.


%
\section{Implementation}\label{S-implementation}

Unlike the standard space-time cloak \cite{McCall-FKB-2011jo}
 and related T-devices \cite{Kinsler-M-2014adp-scast}, 
 these space-time carpet T-devices cannot easily be simplified by
 choosing a preferred direction.
Although they also use speed control of the illumination
 to operate, 
 a significant complication is that a carpet device
 is unavoidably bi-directional
 due to the reflection at the ground plane.
Each location within the cloak halo
 needs to set the speed of incoming illumination
 and outgoing illumination differently; 
 as can be seen in fig. \ref{fig-carpet-Tray}.
This makes it more demanding to implement than
 either a purely spatial carpet device, 
 or an ordinary space-time cloak.

Conceptually, 
 the necessary properties could be achieved
 by making the space-time carpet T-device using a moving medium, 
 but that would require material speeds comparable to the wave speed.
Given the problems of sourcing and sinking the medium
 at its moving boundaries,
 this is unlikely to be straightforward; 
 although we presume that it would be less difficult 
 for water waves than for EM.
Because of this, 
 the more obvious approach would seem to be to design
 a switchable metamaterial system.

\begin{figure}
\includegraphics[angle=-0,width=0.92\columnwidth]{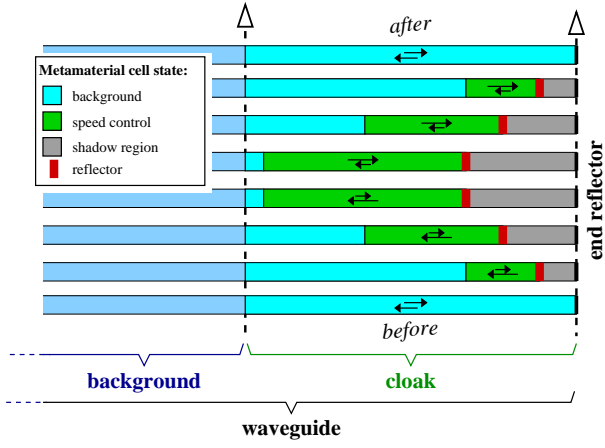}
\caption{(colour online)
A linear (1+1D) waveguide acting as a space-time carpet cloak.
Each horizontal bar represents the waveguide at a moment in time, 
 with early times at the bottom and later times at the top.
The changing state of the waveguide's metamaterial cells
 is indicated by the different colours (shades), 
 with the small arrows indicating the direction sensitive 
 speed contrast. 
}
\label{fig-1dstcarpet}
\end{figure}

As a simple case, 
 consider a space-time carpet cloak added
 to the end of a 1D waveguide terminated by a reflector; 
 as depicted in fig. \ref{fig-1dstcarpet}.
If we break up the cloaking segment of the waveguide
 into a chain of metamaterial cells,
 we find that each cell will need to support either three or four states:

\begin{enumerate}

\item 
 a state that mimics or matches the properties of the rest of the waveguide, 
 i.e. the ``background''.

\item 
 a state that has a direction-dependent speed
 as defined by the cloak design, 
 with incoming illumination slowed
 and outgoing illumination sped up.

\item
 a complementary state with the reverse direction-dependent speeds
 to state 2.

\item
 a state matching the behaviour of the end of the waveguide, 
 which in most cases would be imagined to be a reflector. 
 This capability is only needed 
 for cells that will at some stage form part of the inner boundary
 of the cloaked region.
 
\end{enumerate}

For an EM carpet, 
 magneto-electric effects 
 are required in the design presented in section \ref{S-stcarpet}.
However this means of achieving direction-dependent speeds
 is usually hard to engineer; 
 even having relaxed the requirement for impedance matching.
As an alternative,
 we could distinguish ingoing from outgoing light by 
 altering its frequency
 by some convenient mechanism (as for e.g. \cite{Fridman-FOG-2012nat}), 
 or by using the polarization.
In a 1D waveguide carpet we could control
 ingoing light by making it plane polarized
 in $y$ (hence controlling speed using $\epsilon_y$), 
 and with the help of an engineered polarization switching reflection
 \cite{Zhu-FZHWJ-2010oe}, 
 control the outgoing speed with $z$ polarization and $\epsilon_z$.
The necessary polarization conversion at the entry and exit of the carpet 
 can be applied using a Faraday rotator or comparable device.
However, 
 using plane polarized light means that the reflective state of those
 carpet cells needs to also be polarization switching.
Since the engineering of a four-state metamaterial cell
 is already a challenge, 
 we would prefer to avoid making one of those states more complicated
 than necessary.

Therefore we could instead use chiral metamaterial cells 
 with incoming light distinguished by being 
 (e.g.) RH circularly polarized, 
 and outgoing light LH circularly polarized.
This scheme means that 
 an ordinary mirror-like cell state is sufficient
 to reflect the outgoing light into the correct (opposite) 
  circular polarization.
A further advantage is that it can be implemented
 as a purely dielectric device, 
 so avoiding any need to engineer a controllable magnetic response.

\begin{figure}
\includegraphics[angle=-0,width=0.32\columnwidth]{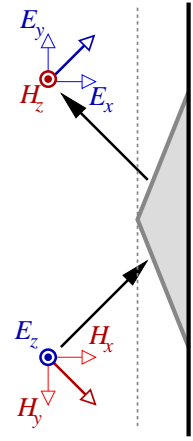}
\caption{
Incoming and outgoing fields 
 for a 1+2D EM space-time carpet cloak implementation; 
 but drawn as if for a spatial carpet cloak for clarity.
In the space-time carpet of section \ref{S-stcarpet}, 
  the reflective carpet pushes out to $x=-L$ (dashed line)
  and back for all $y$ equally.
Incoming fields are controlled by $\epsilon_z$ acting on $E_z$, 
 while 
 the outgoing fields are controlled
 by $\mu_z$ acting on $H_z$; 
 the field polarization is switched on reflection from the carpet.
}
\label{fig-bibicarpet}
\end{figure}

For an all-angle 2D space-time carpet T-device, 
 the implementation gets harder --
 we can no longer rely on using circularly polarized light and a chiral medium.
However, 
 if we revert to linear polarization, 
 and manage to implement the polarization-switching reflector state, 
 it can be achieved by means of a kind of novel birefringence 
 that creates the speed contrast
 between incoming and outgoing illumination.
Here, 
 as depicted in fig. \ref{fig-bibicarpet}, 
 we control the incoming illumination by making
 the electric field component $z$-polarized
 and modulating the speed using $\epsilon_z$.
Since the magnetic field component is then in the $xy$ plane, 
 we set $\mu_x$ and $\mu_y$ to a background value,
 so that the inward speed will remain angle-independent.
Next, 
 the outgoing illumination is controlled by making the 
 magnetic field component $z$-polarized
 and modulating the speed using $\mu_z$.
Since the electric field component is then in the $xy$ plane, 
 we set $\epsilon_x$ and $\epsilon_y$ to a background value,
 so that the outward speed will remain angle-independent.

Whilst this will certainly be 
 challenging to construct,
 one important simplification 
 is that the dynamic $\epsilon_z$ and $\mu_z$ properties required
 are matched in strength to each other, 
 which will hopefully simplify the necessary technology.

In p-acoustics, 
 the scalar nature of the wave's pressure component
 prohibits us from accessing a similar trick, 
 and so there is no alternative but to consider how we might
 achieve the unconventional coupling
 where the velocity field would be able to cause
 changes to the pressure.
Once again, 
 although the general nature of the mathematical formalism 
 has enabled us to unify the transformation aspects of our chosen T-device, 
 the specific physics of a given type of wave
 nevertheless restricts what can be done in principle, 
 as well as in practice.

%
\section{Summary}\label{S-summary}

We have derived the material parameters needed
 for new kind of event cloak --
 the space-time carpet cloak --
 and shown what material properties are required to implement it.
By unifying the mathematical treatment of EM and a scalar wave theory
 such as pressure acoustics, 
 we have also shown the way towards a means of unifying 
 transformation theories for different wave types.
The unified transformation mechanics  theory was applied
 to a range of spatial T-devices based on the ordinary carpet cloak, 
 e.g. the {\peephole} and the tardis.
As noted in the original event cloak paper \cite{McCall-FKB-2011jo}, 
 a covariant wave notation greatly facilitates 
 the design of space-time T-devices \cite{Kinsler-M-2014adp-scast}, 
 which is further emphasized here,
 with a 
 scalar wave theory being 
 matched up with the usual covariant formulation of EM, 
 and the subsequent ease of  
 comparison and contrast 
 between the spatial carpet T-devices 
 from section \ref{S-carpet},
 with the space-time carpet T-devices from section \ref{S-stcarpet}.
We also proposed (in section \ref{S-implementation})
 a scheme laying out the required metamaterial properties
 needed when implementing 
 an electromagnetic space-time carpet T-device.

%

\begin{acknowledgments}
  We acknowledge financial support from EPSRC 
  grant number EP/K003305/1.
\end{acknowledgments}

%
\bibliography{/home/physics/_work/bibtex.bib}

%
\end{document}